\documentclass{article}

\usepackage[english]{babel}
\usepackage{amsmath}
\usepackage{amsfonts}
\usepackage{mathtools}
\usepackage{bbm}
\usepackage{natbib}
\usepackage{xfrac}

\usepackage[letterpaper,top=2cm,bottom=2cm,left=3cm,right=3cm,marginparwidth=1.75cm]{geometry}

\usepackage{amsmath}
\usepackage{graphicx}
\usepackage{xcolor}
\usepackage[colorlinks=true, allcolors=blue]{hyperref}

\usepackage[standard]{ntheorem}
\theoremheaderfont{\normalfont\itshape}

\theoremsymbol{\ensuremath{\blacksquare}} 

\theoremstyle{nonumberplain}

\title{Restricted mean survival time estimation using covariate adjusted pseudovalue regression to improve precision}
\author{Yunfan Li, Jessica L. Ross, Aaron M. Smith, David P. Miller, \\the Pooled Resource Open-Access ALS Clinical Trials Consortium\footnote{Data used in the preparation of this article were obtained from the Pooled Resource Open-Access ALS Clinical Trials (PRO-ACT) Database. As such, the following organizations and individuals within the PRO-ACT Consortium contributed to the design and implementation of the PRO-ACT Database and/or provided data, but did not participate in the analysis of the data or the writing of this report:
ALS Therapy Alliance,
Cytokinetics, Inc.,
Amylyx Pharmaceuticals, Inc.,
Knopp Biosciences,
Neuraltus Pharmaceuticals, Inc.,
Neurological Clinical Research Institute, MGH,
Northeast ALS Consortium,
Novartis,
Prize4Life Israel,
Regeneron Pharmaceuticals, Inc.,
Sanofi,
Teva Pharmaceutical Industries, Ltd.,
The ALS Association.}}

\begin{document}
\maketitle

\begin{abstract}
Covariate adjustment is desired by both practitioners and regulators of randomized clinical trials because it improves precision for estimating treatment effects. However, covariate adjustment presents a particular challenge in time-to-event analysis. We propose to apply covariate adjusted pseudovalue regression to estimate between-treatment difference in restricted mean survival times (RMST). Our proposed method incorporates a prognostic covariate to increase precision of treatment effect estimate, maintaining strict type I error control without introducing bias. In addition, the amount of increase in precision can be quantified and taken into account in sample size calculation at the study design stage. Consequently, our proposed method provides the ability to design smaller randomized studies at no expense to statistical power. [Keywords: clinical trials, covariate adjustment, precision, restricted mean survival time, statistical power, survival analysis]
\end{abstract} \hspace{10pt}

\section{Introduction}
\label{sec:intro}

Survival analysis, also known as time-to-event analysis, is commonly used in randomized controlled clinical trials (RCTs) to assess the treatment benefit of investigational drugs over the length of the study. Various statistical methods have been applied to survival analysis over decades, yet researchers are still debating whether the underlying proportional hazard assumption in one of the most commonly used methods is met in clinical trials \citep{rahman2018violations}. Selecting an estimand and interpretation of the estimand are also open questions in survival analysis \citep{uno2014moving, rufibach2019treatment}. In addition, recent guidelines from the FDA strongly encourage improving the precision of estimand prediction in RCTs via covariate adjustment \citep{food_and_drug_administration_adjusting_2021}, but covariate adjustment is difficult to implement in survival analysis due to the following reasons.

In some widely accepted methods in survival analysis, such as Kaplan-Meier (KM) estimation and the log-rank test, covariates are usually incorporated as stratification factors, which precludes the use of many covariates and results in loss of information with continuous covariates. The Cox regression model takes continuous covariates, but its proportional hazard assumption cannot hold both with and without the covariate, so a correctly specified model is key to inference \citep{schmoor1997effects}. Additionally, the hazard ratio in the Cox regression model is generally non-collapsible \citep{food_and_drug_administration_adjusting_2021}, meaning that the marginal treatment effect differs from subgroup-specific conditional treatment effects, making it difficult to interpret if covariate adjusted. Efforts have been made to address these difficulties. For instance, \citep{lu2008improving}  provides a covariate auxiliary test of the log hazard ratio that doesn't suffer from non-collapsibility. However, its efficiency gain over non-adjusted log hazard ratio is very difficult to predict, and the estimand relies on the proportional hazard assumption as the Cox regression model does. These factors ultimately prevent the widespread adoption of covariate adjustment for precision gain in survival analysis.

In recent years, restricted mean survival time (RMST) has been growing in popularity \citep{nemes2020, perego2020utility, trinquart2016}. RMST does not require a proportional hazards assumption. Parametric models can be built to estimate RMST, which allows for the inclusion of continuous covariates in the model, and the model doesn't have to be correctly specified for inference \citep{tian2014predicting, uno2014moving}. The interpretation of treatment effect from RMST is much more straightforward---a collapsible difference between two means \citep{royston2013restricted, pak2017interpretability}, so that RMST fits the estimand framework and has a natural interpretation.

In this paper, we are interested in the application of covariate adjustment in RMST estimation to improve precision, and particularly in the context of RCTs. We will first examine the parametric estimation of RMST via pseudovalue regression and its underlying statistical properties. Then, we will present an expression to
show that covariate adjusted estimation of RMST significantly improves upon the non-parametric KM-based RMST by reducing variance of the estimated treatment effect, regardless whether covariates are fixed or random. Next, we will show via a number of simulation studies of different types that covariate adjustment increases the precision of RMST difference estimation. Finally, this method will be applied to a case study of RCT data to validate the applicability of covariate adjusted pseudovalue regression. Altogether, this novel survival analysis method improves upon previous methodology in that it both adheres to the guidelines of the major regulatory bodies and increases statistical power without bias and while still maintaining strict Type I error control.

\section{Pseudovalue regression for RMST}
\label{sec:RMST}

\subsection{RMST in survival analysis}
\label{subsec:RMST}

Consider a survival study on the life time $Y^{\circ}$ of subjects. Let $C$ refer to the censoring time, and $C$ may or may not be random relative to $Y^{\circ}$. Then the observed lifetime is $Y=\mathrm{min}(Y^{\circ},C)$. A non-parametric approach is usually taken to estimate $F^{\circ}$, the c.d.f. of $Y^{\circ}$, or equivalently $S^{\circ}$, the survival function of $Y^{\circ}$, where $S^{\circ}(t)=1-F^{\circ}(t)$. The most widely accepted procedure for estimating $S^{\circ}$ is the KM method.

Interpreting differences in distributions is not easy, so it is of some interest to compare functions of estimated survival distributions $\hat{S}$ in randomized studies. In some cases, the mean survival time by KM estimation, or $\int_{0}^{\infty} \hat{S}_{KM} \mathrm{d}t$, may be a relevant metric. However, the experimenter is almost never able to observe the $Y$ of every subject because the study may end while some subjects remain at risk. In addition, quality of the KM estimate $\hat{S}$ is less reliable toward the tail of the survival curve, when data from fewer subjects are available due to censoring.

In recent years, RMST, defined as $\int_0^{\tau} \hat{S} \mathrm{d}t$ with restriction time $\tau$, has grown in popularity due to its ease of interpretation \citep{nemes2020, perego2020utility, trinquart2016, royston2013restricted}. RMST estimates can be obtained using non-parametric KM estimated survival functions. However, this method prevents covariate inclusion, limiting its use. Alternatively, parametric regression models can be utilized to estimate RMST. Two of these methods are the pseudovalue regression method \citep{andersen2003generalised} and the inverse probability censoring weighting (IPCW) method \citep{tian2014predicting}. The pseudovalue regression method assumes that censoring is non-informative to the survival outcomes, and treats the censoring distribution as a nuisance in the estimation. In addition to the non-informative censoring assumption, the IPCW makes the further assumption that the censoring distribution can be properly estimated and weights these values in the estimation process. This additional assumption is impossible to verify in application, and the added complexity of the model makes its variance more difficult to study. In this article, we will focus on pseudovalue regression.

\subsection{Pseudovalue regression}
\label{subsec:pseudovalue_reg}

The pseudovalue regression fits a generalized linear model to time-to-event data. Let $Y_1, ..., Y_n$ be i.i.d. quantities and let $\theta=E[f(Y)]$ for some function $f(.)$, where $\theta$ could be expected survival rate, mean survival time, or RMST. For the purposes of this article, we are only considering RMST as $\theta$. Let $x_1, ..., x_n$ be i.i.d. samples of predictors, and $\theta_i=E[f(Y_i)|x_i]$ be the conditional expectation of $f(Y_i)$ given $x_i$. Direct estimation of $\theta$ with observed survival times is not possible due to censored data. Instead, use an unbiased estimator $\hat{\theta}$ of $\theta$, and define the $i^{\rm th}$ pseudo-observation of $\theta$ as
\begin{equation*}
    \hat{\theta}_i = n\hat{\theta}-(n-1)\hat{\theta}^{-i},
\end{equation*}
where $\hat{\theta}^{-i}$ is the jackknife leave-one-out estimator of $\theta$ based on $\{Y_j: j\neq i\}$, using the same unbiased estimating method.

\citep{andersen2003generalised} proposed to use a generalized linear model $g(\theta_i)=x_i'\beta$ to model the effect of predictors on RMST, and the pseudovalue regression solves the following estimating equation
\begin{equation}
\label{eq:1}
    U(\beta)=\sum_i U_i(\beta)=\sum_i (\frac{\partial \theta_i}{\partial \beta})'V_i^{-1}(\hat{\theta}_i-\theta_i)=0,
\end{equation}
where $V_i$ is a working covariance matrix, often chosen to be the identity matrix.  This framework is very flexible, since the choices of $\theta$ and $g()$ provide various choices of estimands.

In the current paper, we only consider RMST as $\theta$, and the identity link as $g()$. In this case, the model is on restricted survival time, hazard does not play any role in the model, and therefore the proportional hazard assumption is not needed. Sec \ref{subsec:ols} will present properties of the pseudovalue regression, so that our main result on variance reduction in Section \ref{sec:var_red} can be introduced based on these properties.

\subsection{An ordinary least squares estimate}
\label{subsec:ols}

Suppose that in the estimating equation, we choose the identity function as the link function so that $g(\theta)=\theta$, and the identity matrix as $V_i$, then the pseudovalue regression solves $\beta$ to minimize the sum of squares $\sum_i(\hat{\theta}_i-x_i' \beta)^2$. In other words, the pseudovalue regression model becomes a linear regression model instead of a generalized regression model. In addition, the estimating function (\ref{eq:1}) minimizes the derivative of sum of squared residuals. In other words, the pseudovalue regression gives the ordinary least squares (OLS) estimate just like a linear regression does, except that the observed data Y are replaced by pseudovalues to accommodate censoring.

Since the objective function in the pseudovalue regression is a sample average, the method can also be considered as an M-estimator. Therefore, the asymptotic theory of M-estimator applies. In the article that proposed pseudovalue regression, \citep{andersen2003generalised} raised two conditions to establish asymptotic properties of the estimate $\hat{\beta}$: $E[U(\beta^{\circ})]=0$ where $\beta^{\circ}$ is the true parameter, and $U_i(\beta)$ for $i=1,...,n$ are independent. The first condition is satisfied as long as $\hat{\theta}$ is an unbiased estimator of $\theta$. The KM estimate is unbiased except for in the tail \citep{meier_1975}, so unbiasedness is satisfied in RMST as long as restriction time is not too late so that unbiasedness of KM estimate is a concern. In addition, \citep{andersen2003generalised} considered $U_i({\beta})$ for $i=1,...,n$ to be independent in practice, and established asymptotic properties of $\hat{\beta}$. That is, $n^{1/2}(\hat{\beta}-\beta^{\circ})$ is asymptotically normal with mean zero and a variance-covariance matrix estimated by the sandwich estimator $\hat{\Sigma}=A^{-1}BA^{-1}{'}$, where
\begin{equation*}
    A = E[-\frac{\partial}{\partial\beta} U_i(\beta^{\circ})], \quad
    B = E[U_i(\beta^{\circ})U_i(\beta^{\circ})'].
\end{equation*}
However, \citep{jacobsen2016note} showed that the sandwich estimator above is not consistent, as \citep{andersen2003generalised} failed to consider a lower order term that does not vanish asymptotically. As a result, the estimator above results in overestimation of variance, leading to conservative tests. Nevertheless, the numerical studies in \citep{jacobsen2016note} show that the bias is small in reasonable simulation settings. Thus, we will continue to use the sandwich estimator for asymptotic variance of $\hat{\beta}$, and show in simulations that coverage of 95\% confidence intervals generated using the sandwich estimator is not overly conservative.

\section{Variance reduction by covariate adjustment in pseudovalue regression estimated RMST}
\label{sec:var_red}

In the section above, we reviewed pseudovalue regression on RMST and how it can be viewed as an OLS estimator. We further have a sandwich estimator for the variance-covariance matrix of the regression parameters. As a result, we have the necessary ingredients to study how covariate adjustment increases efficiency in pseudovalue regression. In this section, we will first consider correctly specified covariate adjusted models in Sec 3.1, and then incorrectly specified models in Sec 3.2, and give formulas for the performance of pseudovalue regression in each case. We will then connect the efficiency of the pseudovalue regression to that of the KM estimated RMST. Finally, we will discuss how machine learning methods can be utilized to generate a covariate at baseline to improve efficiency in survival analysis.

\subsection{Correctly specified covariate adjusted model}
\label{subsec:fixedx}

First we consider a pseudovalue regression with a correctly specified covariate and study the influence of the covariate on the precision of treatment effect estimation. Suppose that the predictors in the pseudovalue regression $x_i'=(1,\mathbbm{1}_{Ti},u_i)$, corresponding to an intercept, the treatment assignment, and a centered continuous covariate with mean $0$. Here, we first consider the case where the covariate is correctly specified. An example of a correctly specified covariate could be a baseline measure of a demographic characteristic or a biomarker measured without error, believed to be prognostic. Using pseudovalue regression given in Section \ref{subsec:ols}, the estimating equation is 
\begin{equation*}
    \Sigma_{i=1}^n U_i(\beta)=\Sigma_{i=1}^n x_i'(\hat{\theta}_i-x_i'\beta),
\end{equation*}
solving for a least squares estimate of the regression
\begin{equation*}
    \hat{\theta}_i=\beta_0+\beta_1\mathbbm{1}_{Ti}+\beta_2 u_i+\epsilon_i.
\end{equation*}
We will derive the sandwich estimator of the variance-covariance matrix, and then compare the variance of the covariate adjusted pseudovalue regression estimated $\hat{\beta}_1$ (corresponding to the treatment effect) to the unadjusted one.

Denote the percentage of subjects in the treatment arm to be $\pi$, and let the `residual' in the pseudovalue regression be $\epsilon_i=\hat{\theta}_i-x_i'\hat{\beta}$. Since $E[\hat{\theta}]=\theta$, estimate $\hat{\beta}$ is unbiased from the true parameter $\beta^{\circ}$. Therefore $E[\epsilon]=0$.

Recall that the sandwich estimator for the variance-covariance of $\hat{\beta}$ is $A^{-1}BA^{-1}{'}/n$, and
\begin{align*}
    A &= E[-\frac{\partial}{\partial\beta} U_i(\beta^{\circ})] 
    = E \begin{bmatrix}
        1 & \mathbbm{1}_{Ti} & u_i \\
        \mathbbm{1}_{Ti} & \mathbbm{1}_{Ti} & \mathbbm{1}_{Ti}u_i \\
        u_i & \mathbbm{1}_{Ti}u_i & u_i^2
    \end{bmatrix} 
    =\begin{bmatrix}
        1 & \pi & \mu_u \\
        \pi & \pi & \pi \mu_u \\
        \mu_u & \pi \mu_u & \sigma_u^2
    \end{bmatrix}, \quad \mathrm{and} \\
    B &= E[U_i(\beta^{\circ})U_i(\beta^{\circ})']
    = E \begin{bmatrix}
        \epsilon_i^2 & \mathbbm{1}_{Ti}\epsilon_i^2 & u_i\epsilon_i^2 \\
        \mathbbm{1}_{Ti}\epsilon_i^2 & \mathbbm{1}_{Ti}^2\epsilon_i^2 & u_i\epsilon_i^2 \\
        u_i\epsilon_i^2 & \mathbbm{1}_{Ti}u_i\epsilon_i^2 & u_i^2\epsilon_i^2
    \end{bmatrix}
    =\begin{bmatrix}
        \sigma_{\epsilon}^2 & \pi\sigma_{\epsilon}^2 & \mu_u\sigma_{\epsilon}^2 \\
        \pi\sigma_{\epsilon}^2 & \pi\sigma_{\epsilon}^2 & \mu_u\sigma_{\epsilon}^2 \\
        \mu_u\sigma_{\epsilon}^2 & \pi\mu_u\sigma_{\epsilon}^2 & \sigma_u^2\sigma_{\epsilon}^2
    \end{bmatrix},
\end{align*}
since residuals $\epsilon$ have mean zero and are orthogonal to the predictors. Here, $\mu_u$ and $\sigma_u^2$ denote the expectation and variance of the covariate, and $\sigma_{\epsilon}^2$ denotes the residual variance.

Some calculations show that the variance of $\hat{\beta}_1$ (corresponding to the treatment effect) is $\sigma_{\epsilon}^2/(n\pi(1-\pi))$. Recall that $\sigma_{\epsilon}^2$ in the covariate adjusted model is smaller than the unadjusted variance, $\mathrm{Var}(\hat{\theta}_i)$. To be precise, $\sigma_{\epsilon}^2=\mathrm{Var}(\epsilon)=(1-r^2)\mathrm{Var}(\hat{\theta}_i)$, where $r$ is the correlation between $\hat{\theta}_i$ and covariate $u_i$, so that $r^2$ describes the percentage of variance in $\hat{\theta}_i$ that can be explained by $u_i$. When there is a treatment effect and the correlation in the two arms differs, then a weighted correlation should be used
\begin{equation}
\label{eq:r2}
    \sigma_{\epsilon}^2 = (1-\{(1-\pi)r_1+\pi r_0\}^2)\mathrm{Var}(\hat{\theta}_i),
\end{equation}
where $r_0$ is the correlation between pseudovalues and covariates in the control arm, and $r_1$ is the correlation between pseudovalues and covariates in the treatment arm. With this estimated variance $\sigma_{\epsilon}^2/(n\pi(1-\pi))$, a confidence interval of $\hat{\beta}_1$ can be generated, and a statistical test with a chosen type I error can be conducted.

\subsection{Incorrectly specified covariate model}
\label{subsec:randomx}
In Sec \ref{subsec:fixedx}, we considered covariate adjustment in correctly specified models and gave the formula for variance reduction. However, not all models can be considered correctly specified. When it comes to model specification, whether the treatment truly has a linear effect on RMST does not matter. As discussed in \citep{imbens}, causal effect is estimated correctly by linear regression in randomized studies regardless of whether the linear form holds. For covariate adjustment, the specification of the covariates is more important, as incorrectly specified covariate(s) lead to either under- or over-confidence in variance reduction and precision gain.

Covariates may have random error or systematic bias, as they may be measured with error or they may be estimated from an external model. The existence of systematic bias would not affect the performance of pseudovalue regression, as any bias will be absorbed by the intercept term in the linear model. To study the effect of random errors in covariates, we consider, without loss of generality, a single covariate with random errors.

In this situation, we are adopting the framework of data uncertainties in least squares regression by \citep{hodges1972data} and \citep{davies1975effect}. Here, the regression model still has the form $\hat{\theta}_i=\beta_0+\beta_1\mathbbm{1}_{Ti}+\beta_2 u_i+\epsilon_i$. However, we replace $u_i$ with the composition $c_i=u_i+\delta_i$, in which $\delta$ represents some ($\epsilon$-independent) measurement error added to the true, unknown covariate $u_i$.  An example would be an external model that generates estimates $c_i$ of each subject's unknown prognostic score $u_i$ based on the subjects' baseline characteristics. As all estimates are imperfect, the estimates $c_i$ can be viewed as having some random error $\delta_i$.  Let $E[\delta]=0$, and $\mathrm{Var}(\delta)=\sigma^2_{\delta}$.  The former can be assumed without loss of generality, because any bias in $\delta$ will be absorbed by the intercept $\beta_0$, and will not affect the estimate of the treatment effect $\beta_1$ as discussed above.

In this setting, the least squares estimate $\hat{\beta}_{\Delta}$ is generated using the observed covariates $c_i$ instead of using the true but unmeasurable covariates $u_i$, so $\hat{\beta}_{\Delta}$ is no longer unbiased. Let $E[\hat{\beta}_{\Delta i}]=\beta^{\circ}_i-b_i$ so $b_i$ is the bias in the $i$th coefficient. Applying Theorem 4.1 in \citep{davies1975effect}, as $n \to \infty$, the bias of the treatment effect estimate $\hat{\beta}_{\Delta 1}$ converges
\begin{align*}
    b_1 \xrightarrow[]{p} \frac{\beta_2^{\circ}}{\frac{\mathrm{Var}[u]/\sigma^2_{\delta}+1}{\sqrt{\mathrm{Var}[u]/(\pi(1-\pi)n)}Z}-\sqrt{\mathrm{Var}[u]\pi(1-\pi)/n}\frac{Z}{\sigma^2_{\delta}}},
\end{align*}
where $Z$ is a standard normal distributed random variable and $\beta_2^{\circ}$ is the true regression coefficient corresponding to the true and unmeasurable covariate $u$. Although $\mathrm{Var}[u]$ is not known in practice and this rescaling is not realistic in application, let us rescale both the measured covariate $c$ and unmeasurable covariate $u$ by dividing $\sqrt{\mathrm{Var}[u]}$ for a better understanding of the bias. The ratio $\mathrm{Var}[u]/\sigma^2_{\delta}$ will be the signal-to-noise ratio in the covariate. Let the rescaled $\beta_2^{\circ}$ be denoted by $\tilde{\beta}_2^{\circ}$, and $\tilde{b}_1$ be the bias of the treatment effect after rescaling. Then the bias converges
\begin{align} \label{eq:3}
    \tilde{b}_1 \xrightarrow[]{p} \frac{\tilde{\beta}_2^{\circ}}{\frac{\sqrt{n\pi(1-\pi)}(1+\mathrm{Var}[u]/\sigma^2_{\delta})}{Z}-\sqrt{\frac{\pi(1-\pi)}{n}}\frac{\mathrm{Var}[u]}{\sigma^2_{\delta}}Z}.
\end{align}

After inspecting the above, it can be seen that a larger signal-to-noise ratio $\mathrm{Var}[u]/\sigma^2_{\delta}$ gives a smaller bias. The bias also converges to $0$ if $\mathrm{Var}[u]/\sigma^2_{\delta} \to \infty$, or $\sigma^2_{\delta} \to 0$. Furthermore, we have the following theorem.
\begin{theorem} \label{claim:2}
    If $\mathrm{Var}[u]/\sigma^2_{\delta}$ is bounded, as $n \to \infty$, $\tilde{b}_1 \xrightarrow[]{D} 0$.
\end{theorem}

Theorem \ref{claim:2} states that even if the error $\sigma^2_{\delta}$ does not disappear, then as long as the error is bounded and the sample size in the randomized study is large, the bias in the treatment effect estimate is still asymptotically $0$. This is consistent with the statement in \citep{davies1975effect} that if errors in the columns of the design matrix are independent (in our case, only the column of covariate has errors) and the design is balanced, then the contamination in the coefficient estimate of one component of $\beta$ (in our case, the treatment effect) by the other components (in our case, the covariate) will not be important. A derivation of Theorem \ref{claim:2} is given in Appendix \ref{sec:A1}.

Now that there is satisfactory result on the asymptotic bias, let us consider the variance of the treatment effect estimate. Applying Theorem 4.2 from \citep{davies1975effect}, we have the following results.
\begin{theorem} \label{claim:3}
    As $n \to \infty$, $\mathrm{Var}[\hat{\beta}_{\Delta 1}] \xrightarrow[]{D} \frac{\sigma^2_{\epsilon}+\tilde{\beta}^{\circ 2}_2 \sigma^2_{\delta}/\mathrm{Var}[u]}{n\pi(1-\pi)}$.
\end{theorem}
\begin{theorem} \label{claim:4}
    If $\sigma^2_{\delta} \to 0$ as $n \to \infty$, $\mathrm{Var}[\hat{\beta}_{\Delta 1}] \xrightarrow[]{D} \frac{\sigma^2_{\epsilon}}{n\pi(1-\pi)}$.
\end{theorem}

Theorem \ref{claim:3} states that when sample size in the randomized study is large, variance of the treatment effect is almost variance in the case with correctly specified models, but with an inflation inversely proportional to the signal-to-noise ratio $\mathrm{Var}[u]/\sigma^2_{\delta}$. For this inflation to disappear and the variance with a random covariate to be as good as variance in correctly specified models, the error $\sigma^2_{\delta}$ has to go to $0$ as well, as described in Theorem \ref{claim:4}. A derivation is given in Appendix \ref{sec:A2}. In terms of computation, the standard error generated in the usual way will be a reasonable estimate of the standard error in the presence of random covariates \citep{hodges1972data}, therefore confidence intervals and statistical tests done in the usual way will be reasonable as well.

This section shows that even if the covariate is misspecified, the treatment effect estimate will still be unbiased as long as sample size in the randomized study is large. For the variance reduction by covariate adjustment to hold, the sample size in the randomized study needs to be large, and the error in the observed covariate needs to be small. We will discuss the implications of these results in Sec \ref{subsec:ml}.

\subsection{Influence function and pseudovalue}
\label{subsec:pv}

Sections \ref{subsec:fixedx} and \ref{subsec:randomx} demonstrate the benefits of covariate adjusted RMST estimate compared to the treatment effect estimated by an unadjusted pseudovalue regression. However, pseudovalue regression is not widely adopted in randomized clinical trials yet. It would be much more interesting to compare covariate adjusted RMST to other, preferably more widely adopted estimates, in survival analysis. In this section, we are going to consider KMbased RMST estimate first, and then briefly discuss other estimands in survival analysis.

As described in Sec \ref{subsec:pseudovalue_reg}, pseudovalues are generated by leave-one-out jackknife estimates. Jackknife has a connection to the theoretical influence function, which is used to study variance of an estimator \citep{hampel1974influence}. Suppose $\theta^{\circ}=f(F^{\circ})$ is a function of the true survival distribution, $\hat{\theta}=f(\hat{F})$ is its KM estimate, then the von Mises expansion gives
\begin{equation*}
    \hat{\theta} = \theta^{\circ}+\int \dot{\psi}(y) \mathrm{d}\hat{F}(y) +o_p(1),
\end{equation*}
where $\dot{\psi}(y)$ is the first order influence function \citep{steinbach1982jackknife}. Empirically, since $\hat{F}$ assigns mass $1/n$ to each sample $Y_i$, $i=1,...,n$, the above expansion becomes
\begin{equation*}
    \hat{\theta} = \theta^{\circ}+\frac{1}{n}\sum_{i=1}^n \dot{\psi}(Y_i)+o_p(1),
\end{equation*}
where $\dot{\psi}(Y_i)$ denotes the first order influence function evaluated at $Y_i$.

The influence function can be seen as a measure of the change in $\theta$ if the mass at $y$ is changed by an infinitesimal amount. For observation $i$, $\hat{\theta}_i$ is exactly the statistic $\hat{\theta}$ modified in the direction given by the leave-one-out estimator $\hat{\theta}^{-i}$. It turns out that the pseudovalues are related to $\dot{\psi}(y)$ in the following way \citep{jacobsen2016note}
\begin{equation*}
    \hat{\theta}_i=\theta^{\circ}+\dot{\psi}(Y_i)+o_p(1).
\end{equation*}

In fact, \citep{reid1979influence} states that `the jackknife and the influence curve approaches lead to the same asymptotic calculation because the jackknife is a finite sample approximation to the influence function.' Obtaining the variance of the empirical influence function is hard, but obtaining the variance of the jackknife estimates is much easier. With a sample size of $n$, the pseudovalues give $n$ realizations of empirical influence function evaluated at $Y_i$. Asymptotically, variance of $\hat{\theta}_i$ is equivalent to variance of $\hat{\theta}$.

This argument also provides a theoretical justification for pseudovalue regression. Although pseudovalue regression was proposed twenty years ago, there has been a lack of interpretation of the pseudovalues. However, by realizing the link between the influence function, the pseudovalues generated by the jackknife approach gain an interpretation rather than being a mere computational convenience.

Now consider treatment effect on RMST estimated in the most straightforward and unadjusted way, by KM method: use KM to estimate RMST in the two treatment arms, and obtain the estimates $\hat{\theta}_{trt}$ and $\hat{\theta}_{control}$, respectively, then take their difference to get the KM-based estimate $\hat{\theta}_{KM} = \hat{\theta}_{trt} - \hat{\theta}_{control}$. When $\mathrm{Var}(\hat{\theta}_{trt}) = \mathrm{Var}(\hat{\theta}_{control})$, the KM-based estimate has variance $\mathrm{Var}(\hat{\theta}_{KM}) = \mathrm{Var}(\hat{\theta})/(n\pi(1-\pi))$, where $\hat{\theta}$ is the KM estimated RMST with the two arms pooled.\footnote{If $\mathrm{Var}(\hat{\theta}_{trt}) \neq \mathrm{Var}(\hat{\theta}_{control})$, they can be pooled to get $\mathrm{Var}(\hat{\theta})$.} Substitute $\mathrm{Var}(\hat{\theta})$ by the asymptotically equivalent $\mathrm{Var}(\hat{\theta}_i)$, and the benefits of covariate adjusted RMST discussed in sections \ref{subsec:fixedx} and \ref{subsec:randomx} directly transfer to benefits over $\hat{\theta}_{KM}$.

Finally, other authors have done extensive research to compare the efficiency of RMST to other estimands in survival analysis, most notably the hazard ratio, under various assumptions \citep{tian2018efficiency, huang2018comparison, quartagno2021restricted}. In short, RMST is competitive in terms of statistical power, even when the proportional hazard assumption holds, and always has a clear clinical interpretation, unlike the hazard ratio when the proportional hazard assumption fails. We will not provide extra work to compare adjusted or unadjusted RMST to other estimands.

\subsection{Variance reduction via a covariate derived from machine learning}
\label{subsec:ml}

The previous sections derive the precise relationship between the variance reduction obtained from covariate adjustment for RMST via pseudovalue regression (compared to the KM-based RMST estimate) in terms of the correlation coefficient between the adjustment covariate $u_i$, or $c_i$, and the pseudovalue $\hat{\theta}_i$. 

In the context of a randomized study, a number of baseline covariates, $\{w^1_i, \dotsc, w^N_i\}$, may be measured for each subject.  Using any, or all, of these covariates in the above covariate adjustment scheme has the potential to reduce the variance of the treatment effect estimate.  However, there is no reason that the adjustment covariates be directly measured.  In fact, one may adjust for a covariate derived from a machine learning model that was built from historical data. This amounts to pre-specifying a function $f$ of the baseline covariates, and defining a synthetic variable $u_i := f(w^1_i, \dotsc, w^N_i)$ from its values.  Using one covariate $u_i$ instead of $N$ covariates $\{w^N_i\}$ in the regression takes less degrees of freedom and is desirable, especially if $N$ is large. The function $f$ may be implemented by any mathematical or computational means, and the \emph{a priori} variance reduction is then cast in terms of the correlation between $\hat{\theta}_i$ and $f(w^1_i, \dotsc, w^N_i)$.  

An important observation is that the greatest gain in variance reduction can be obtained by adjusting for the outcome itself, that is $u_i = \hat{\theta}_i$.  Therefore, a primary approach to the design of a randomized study incorporating a machine-learning-derived covariate is to build a predictive model $P$ of $\hat{\theta}_i$ in terms of the measured baseline covariates, $\{w^N_i\}$.  Such a model provides a function $f_P(w^1_i, \dotsc, w^N_i)$, which can be used as an adjustment covariate.  

One approach to this problem is to fit a survival model $P$ that provides an estimate of the conditional survival distribution $p_P(Y > t \vert w^1_i, \dotsc, w^N_i) \simeq p(Y > t \vert w^1_i, \dotsc, w^N_i)$, or the restricted versions thereof.  Assuming one can compute or estimate expectations, taking 
\begin{equation*}
    f_i := \mathbb{E}\left[ p_P(Y \vert w^1_i, \dotsc, w^N_i)\right],
\end{equation*}
provides an adjustment covariate that approaches the optimal adjustment covariate.

If the prognostic score $f_i$ is close to $\hat{\theta}_i$, that is, the error terms $\epsilon_i$ are small due to intrinsic progression of the disease, then the correlation between $f_i$ and $\hat{\theta}_i$ is strong, and covariate adjustment provides a noteworthy increase in the precision of treatment effect estimate.

Indeed, if we analyze this in the setting of random covariate adjustment (\ref{subsec:randomx}), we are interested in the regression problem $$\hat{\theta}_i=\beta_0+\beta_1\mathbbm{1}_{Ti}+\beta_2 f_i+\epsilon_i$$ 
in which $f_i = f(w^1_i, \dotsc, w^N_i)$ for some measured baseline covariates $\{w^N_i\}$.  We regard $f_i$ as an estimate of the true regression covariate $u_i$ and analyze the setting in terms of the theorems in the previous section by specifying $\delta_i := f_i - u_i$, as an $\epsilon$-independent error.  Recall that we can assume $\delta$ has zero-mean because any bias of f is handled by the intercept.
We recover the conclusion that as long as the target trial has randomized treatment assignment and a large sample size, the treatment effect estimation is asymptotically unbiased, with reasonable estimate of the standard error for statistical testing. In addition, the precision of the treatment effect estimate can still be improved from an unadjusted analysis, as long as the correlation between $u_i$ and $\hat{\theta}_i$ is strong, and the ratio $\mathrm{Var}[u]/\sigma^2_{\delta}$ is large.

\section{Simulation studies}
\label{sec:simulation}

In this section, we are going to first show that covariate adjusted pseudovalue regression estimates have comparable bias and coverage to KM-based estimates of RMST difference. We will then show the percentage of variance reduction as a function of correlation in simulation studies, and how the percentage of variance reduction can be explained by our formulas. In our first example, the covariate will be correctly specified and the data generating model will be linear, matching the pseudovalue regression model assumption. We will also show an example using an incorrectly specified covariate. An example of a nonlinear (quadratic) data generating model can also be found in Appendix \ref{sec:A3}.

In our first example, the continuous covariate $u$ follows an exponential distribution of rate $1$. Latent event times $Y^{\circ}$ follow an exponential distribution of rate $1/(a+0.5\mathbbm{1}_T+3u)$, so that the mean survival time (if there is no censoring) is a linear function of the covariate, and the covariate is a perfect prognostic prediction for a patient in the control arm when $a=0$. The constant $a$ takes the value $0$, $0.5$ or $1$ so that the correlation between the covariate and the response changes, and median hazard ratios range from $0.84$ to $0.89$ across values of $a$. Two censoring patterns are considered: one without censoring, and one with censoring, where the censoring time $C$ follows an exponential distribution with rate $0.1$ independent from the survival distribution. Two restriction times are studied, corresponding to the $0.5$ and $0.35$ percentiles of the distribution of $Y^{\circ}$ in the control arm, so that $50\%$ and $65\%$ of subjects in the control arm remain at risk of the event when there is no censoring. For each scenario, $5000$ datasets were generated with each having $500$ samples and the treatment and control arms equal in size. Pseudovalues were obtained using the \texttt{pseudomean} function from the \texttt{`pseudo'} package in R, and heteroskedastic consistent standard errors were obtained using the \texttt{coeftest} function with the HC1 method in R \citep{R}.


\begin{figure}[ht]
\caption{Variance reduction of pseudovalue regression estimated RMST difference compared to KM-based RMST difference, versus squared correlation between covariate and pseudovalues. The grey line is the 45 degree line. The different symbols denote different restriction times, and the different colors denote different distributions for the censoring time. Variance reduction follows Equation \ref{eq:r2}}
\label{fig:linear}
\includegraphics[width=10cm]{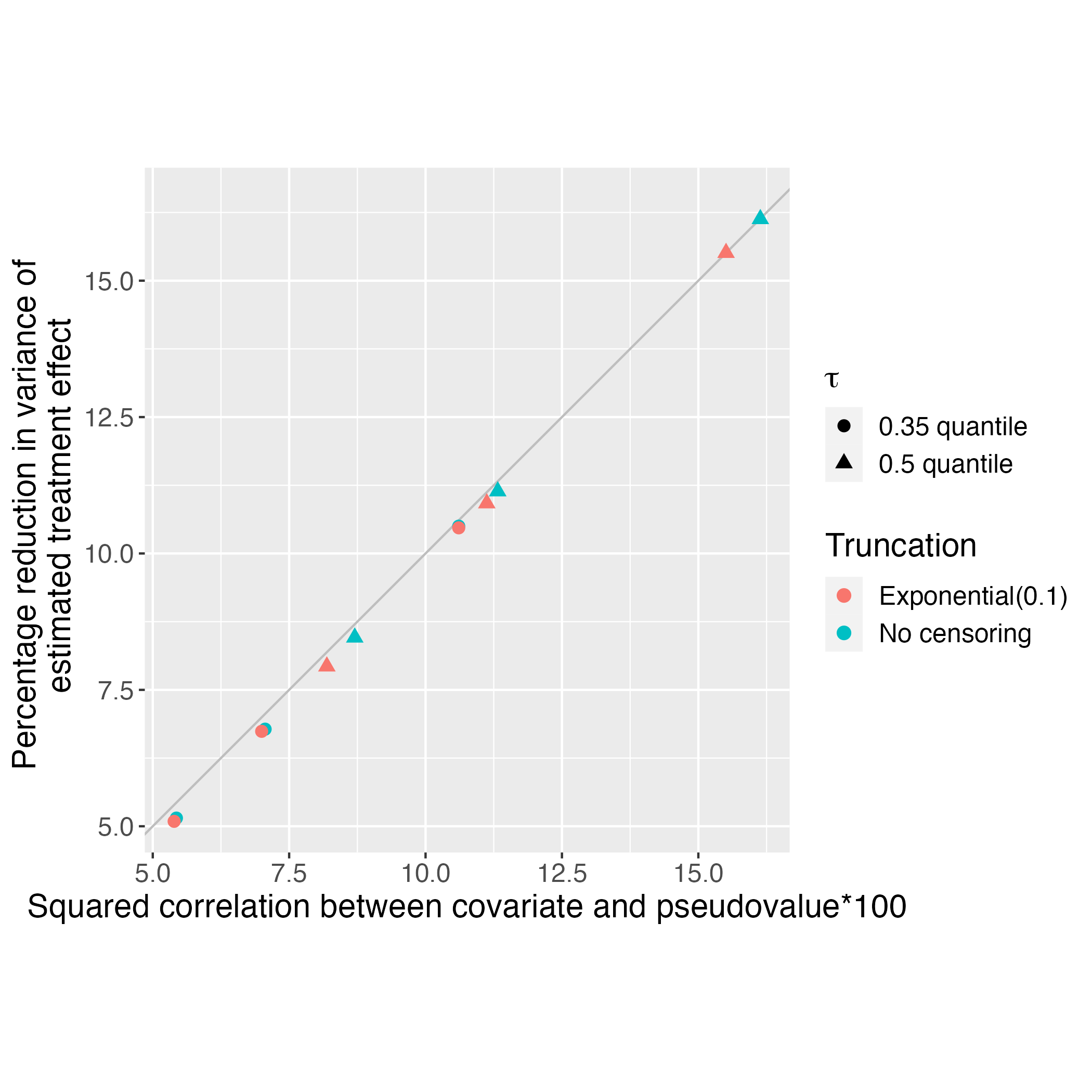}
\centering
\end{figure}

\begin{figure}[ht]
\caption{Variance reduction of pseudovalue regression estimated RMST difference compared to KM-based RMST difference, versus squared correlation between latent covariate (without random error) and pseudovalues. Covariate with random noise. Grey dotted line is the 45 degree line. Grey solid line shows variance reduction versus squared correlation between covariate (with random error) and pseudovalues. The different symbols denote different restriction times, and the different colors denote different distributions for the censoring time. Variance reduction follows Equation \ref{eq:r2} using correlation between observed covariate (with random error) and pseudovalues.}
\label{fig:randomx}
\includegraphics[width=10cm]{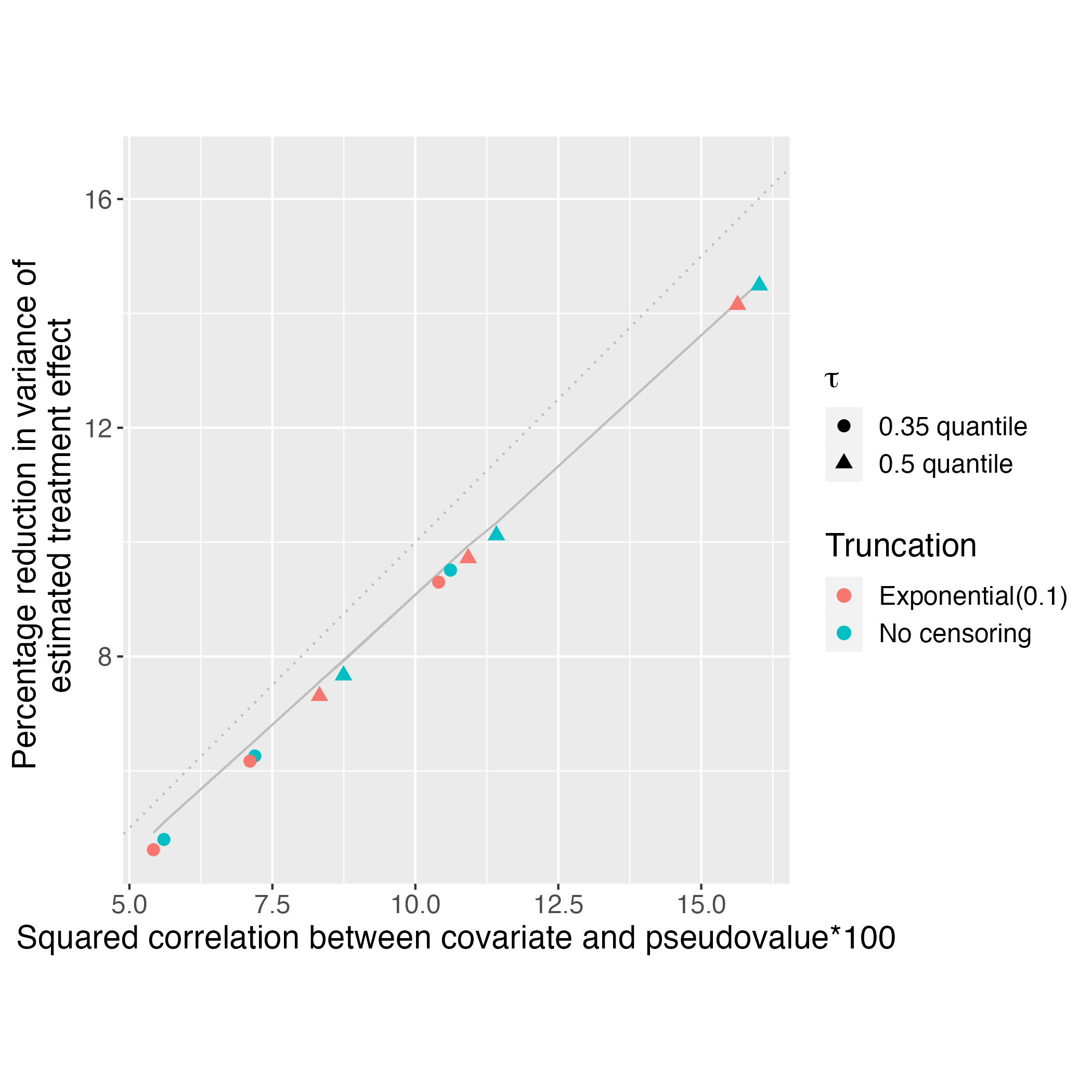}
\centering
\end{figure}

Table \ref{tab:linear} shows covariate adjusted pseudovalue estimated RMST difference compared to KM estimated RMST difference. Expectation of the latent event times is a linear function of the covariate in this table. Pseudovalue regression estimates and KM estimates have comparable bias close to $0$, since both of these estimates are asymptotically unbiased. Both methods have CI coverage close to 95\%, implying that they have the correct type I error rates as well. As expected, pseudovalue regression estimated RMST differences have smaller variance. Here $r^2$ is the squared correlation in the two arms pooled. Since the randomization ratio is $1:1$ in these simulations, the rule-of-thumb $r^2$ variance reduction formula fits well, and we do not use the weighted correlation formula in Equation \ref{eq:r2}. Figure \ref{fig:linear} plots the variance reduction versus squared correlation detailed in Table \ref{tab:linear}. The percentages of variance reduction closely follow the grey line showing squared correlation, as expected from our claim in Section \ref{subsec:fixedx}.

Table \ref{tab:randomx} and Figure \ref{fig:randomx} follow the exact same simulation settings as Table \ref{tab:linear} and Figure \ref{fig:linear}, except that the covariate $c$ is generated with a normal distributed noise with mean $0$ and variance $0.1$ on top of the exponential distributed true covariate $u$. Therefore, the ratio $\mathrm{Var}(u)/\sigma^2_{\delta}$ is $10$ in these simulations. Correlations reported in Table \ref{tab:randomx} and Figure \ref{fig:randomx} are between pseudovalues and the latent true covariate with the two arms pooled. As predicted by our theorems in Section \ref{subsec:randomx}, bias and CI coverage of covariate adjusted pseudovalue regression show almost no impact by the random error in the covariate in these simulations, such that the estimate is still (asymptotically) unbiased with the same type I error control. Compared to expected variance reduction projected by the correlation between the latent true covariate and the pseudovalues (shown by the dotted line in Figure \ref{fig:randomx}), the actual variance reduction is smaller, by a magnitude of $1.5\%$ when correlation is $0.4$, and $0.5\%$ when correlation is $0.25$. Covariate adjustment still improves precision of the estimate compared to the non-adjusted KM-based estimator, since the variance reduction is always above $0$. Note that the actual variance reduction corresponds to the squared correlation between the observed covariate and the pseudovalues (shown by the solid line), which is smaller than the former squared correlation because of errors in the covariate.

\section{Case Study}
\label{sec:case}

Amyotrophic lateral sclerosis (ALS) is a rare and fatal neurodegenerative disease causing progressive decline. We used historical ALS clinical trial participant data from the Pooled Resource Open-Access ALS Clinical Trials (PRO-ACT) database \citep{PROACT}, accessed on April 2, 2021. PRO-ACT contains curated anonymized data from 23 Phase II and III RCTs conducted between 1990 and 2016, as volunteered  by companies and academic research institutions (PRO-ACT Consortium members). The constituent studies of the PRO-ACT data were anonymized to conceal the study membership of the participants. As no overall benefits were detected in the treatment arms of the included trials, the PRO-ACT data includes data from both placebo and treatment arms of ALS clinical trials. Thus, data from both arm assignments were utilized to build the prognostic model.

Data from 9756 trial participants contained in the PRO-ACT database were partitioned into training and evaluation datasets at a 0.7/0.3 ratio. A prognostic model was trained and cross-validated on the training data split using patients' baseline measures as input, and survival as output. A procedure of 5-fold cross-validation was used to select optimal hyper-parameters for the prognostic model. After the optimal model was selected, the evaluation set was used to estimate the correlation between observed restricted survival time pseudovalues and model-predicted restricted survival times. In the evaluation set, the correlation between observed restricted survival time pseudovalues and model-predicted restricted survival times at 365 days is 0.44.

To test our novel survival analysis methodology, we utilized data from a previous Phase 3 double-blind, placebo-controlled trial of ceftriaxone infusion in ALS (NCT00349622). The ceftriaxone study is not contained in the PRO-ACT database and was obtained from the National Institute of Neurological Disorders and Stroke (NINDS). A total of 513 participants were randomized at a 2:1 ratio (2-4 g ceftriaxone : placebo) tracked for 52 weeks following randomization of the last subjects \citep{cudkowicz2014safety}.

The co-primary outcome measures used in the ceftriaxone study were the rate of functional decline as quantified by change in the ALS functional rating scale-revised (ALSFRS-R) over time, and the time to death, tracheostomy, or the start of permanent assisted ventilation, of which the latter served as the primary outcome of interest. For the purpose of a case study in the current paper, only the event of death is considered in survival analysis, and a restriction time is set at 365 days. At the end of 365 days, $18.1\%$ of the 513 subjects perished, $11.9\%$ were censored, and $70.0\%$ remained at risk of the event. For the purpose of this case study, subjects on the two doses of ceftriaxone were combined into a single treatment arm.

Using the \texttt{lifetest} procedure in SAS/STAT 15.1 (SAS Institute, Cary, NC), the KM-based estimate of the treatment effect of ceftriaxone on restricted mean survival time is $-6.88$ days, with a standard error of $5.19$. Using the prespecified prognostic model developed on the PRO-ACT database, a prognostic restricted survival time was generated for each subject. The correlation between the prognostic scores and pseudovalues of outcomes is $0.37$ overall. The correlation is $0.41$ in the control arm and $0.35$ in the treatment arm, respectively. Using the \texttt{rmstreg} procedure in SAS/STAT, prognostic adjusted pseudovalue regression estimates the treatment effect to be $-5.31$ days, with a standard error of $4.76$. Covariate adjusted pseudovalue regression provides a variance reduction of $16.1\%$ from KM estimated treatment effect, close to the $15.4\%$ prediction using weighted correlations in Equation \ref{eq:r2}. A summary of the results are shown in Table \ref{tab:case study}.

\section{Discussion}
\label{sec:discussion}

Randomized clinical trials are time-consuming, resource-intensive, and rely on the consistent participation of patient volunteers for their success. In this paper, we propose to utilize the covariate adjusted pseudovalue regression to increase precision of an unbiased treatment effect estimate while strictly controlling type I error. In randomized clinical trials, this solution can decrease sample sizes without loss of power, and has the potential to reduce both the financial burden and patient burden associated with clinical trials and to bring safe and effective new therapies into clinical practice sooner.

Our method builds upon the RMST framework, which has been gaining popularity for time-to-event analyses due to its collapsibility, interpretability, and lack of reliance on the proportional hazards assumption \citep{perego2020utility, emmerson2021understanding}, properties that are highly favored by regulatory agencies \citep{food_and_drug_administration_adjusting_2021, emaCovar}. In the current paper, we point out that RMST difference estimated by pseudovalue regression has the extra benefit of precision gain by covariate adjustment, especially that the amount of precision gain can be quantified prospectively, a huge advantage in study design.

During the process of revising this article, we are aware that Hattori and Uno \citet{hattori} reached a similar result to our in sample size calculation for an RMST-based test by studying the correlations between the baseline covariates and the martingale residuals. From a mathematical point of view, the Hattori and Uno result is more rigorous, although it depends on hard-to-verify assumptions and is difficult to calculate. On the contrary, our result by studying the correlations between covariates and pseudovalues is much easier to apply in practice.

Leveraging historical data, which is increasingly available for research, provides a new opportunity to reduce sample sizes, but single-arm direct enrichment methods are known to increase bias. Luckily, randomized experiments give unbiased estimates. By integrating a prognostic covariate generated from historical data into a randomized trial, the solution proposed here provides a time-to-event complement to the similarly unbiased PROCOVA approach \citep{schuler2021increasing}, which has received a qualification opinion from the EMA \citep{european_medicines_agency_committee_for_medicinal_products_for_human_use_chmp_draft_2022}.

A central finding, which makes the proposed method accessible to clinical trial practitioners, is that the increase in precision and potential reduction in sample size solely rely on the correlation between the prognostic scores and the pseudovalues. An important limitation of the approach is that the expected correlation is associated with the amount of censoring and the proportion of patients who have not yet had the event at the truncation time. Nonetheless, assumptions about censoring and event rates are essential in any power calculation, so this does not pose an entirely new challenge.

Another limitation of this method is that the sample-size reduction benefit of covariate adjustment depends on how well the prognostic score is estimated and how applicable it is to a new dataset. Machine learning methods provide an opportunity for improved predictions, even in the presence of non-linear effects and complex interactions between variables. For application in clinical studies, it is essential that models be tested in out-of-sample data on a population similar to the population in which the future trial will be conducted. With proper attention to testing, an accurate assessment can be made about the predictive power of prognostic scores vs individual covariates. Importantly, prognostic scores must be estimable from baseline covariates alone to maintain the unbiased properties of randomized trials.

Lastly, it must be emphasized that this method is intended for use in randomized trials. In observational studies, imbalances in covariates in groups are common. Absent randomization, covariate adjustment will reduce bias in estimates of between-group differences, but power may actually decrease as a result. Prospective use for designing randomized clinical trials is the context of use for which the proposed methods have the greatest potential value, and this is the context for which the statistical properties described here apply.

\section{Data Availability}
\label{sec:dataavailability}

Certain data used in the preparation of this article were obtained from the National Institute of Neurological Diseases and Stroke’s Archived Clinical Research data (Title: Clinical Trial Ceftriaxone in Subjects with Amyotrophic Lateral Sclerosis, PI: E. Cudkowicz, MD, Funding: NINDS 5 U01-NS-049640) received from the Archived Clinical Research Dataset web site (https://www.ninds.nih.gov/current-research/research-funded-ninds/clinical-research/archived-clinical-research-datasets).

Other data used in the preparation of this article were obtained from the Pooled Resource Open-Access ALS Clinical Trials (PRO-ACT) Database, where they have been volunteered by members of the PRO-ACT Consortium: ALS Therapy Alliance, Cytokinetics, Inc., Amylyx Pharmaceuticals, Inc., Knopp Biosciences, Neuraltus Pharmaceuticals, Inc., Neurological Clinical Research Institute, MGH, Northeast ALS Consortium, Novartis, Prize4Life Israel, Regeneron Pharmaceuticals, Inc., Sanofi, Teva Pharmaceutical Industries, Ltd., and The ALS Association.

\section*{Acknowledgements}
We wish to thank Eric Tramel, Jason Christopher, Luca D'Alessio, and Alyssa Vanderbeek for their helpful feedback and comments. We also thank Jon Walsh, Daniele Bertolini, Katelyn Arnemann, and Jamie Reiter for their contributions to the case study.

\section*{Financial Disclosure}
YL, JLR, AMS, and DPM are equity-holding employees of Unlearn.ai, Inc., a company that creates software for clinical research and has patents pending for work described (US 17/808,954) and referenced (17/074,364) herein. The PRO-ACT Consortium includes members from the biopharmaceutical industry.

\clearpage
\bibliographystyle{plainnat}
\bibliography{rmst}

\newpage
\begin{table}[h]
    \centering
    \begin{tabular}{c c c c c c c c}
        \hline
        Censored & Truncated & $r^2$ & KM & PV & Variance & KM & PV \\
        (\%) & (\%) &  & bias & bias & reduction (\%) & coverage (\%) & coverage (\%) \\
        \hline
        0.0 & 53.8 & .40 & .020 & .020 & 16.1 & 94.40 & 94.76 \\
        0.0 & 53.3 & .34 & .006 & .006 & 11.1 & 94.68 & 94.40 \\
        0.0 & 52.9 & .30 & .001 & .001 & 8.5 & 94.46 & 94.68 \\
        0.0 & 69.9 & .33 & .011 & .011 & 10.5 & 94.74 & 94.76 \\
        0.0 & 69.0 & .27 & .002 & .002 & 6.8 & 94.56 & 94.84 \\
        0.0 & 68.4 & .23 & -.001 & -.001 & 5.1 & 95.26 & 94.98 \\
        8.1 & 48.1 & .39 & .023 & .023 & 15.5 & 94.74 & 94.70 \\
        11.4 & 44.1 & .33 & .013 & .013 & 10.9 & 94.70 & 94.86 \\
        13.6 & 43.7 & .29 & .000 & .000 & 7.9 & 94.44 & 94.86 \\
        4.8 & 65.4 & .33 & .012 & .012 & 10.5 & 94.90 & 94.54 \\
        7.2 & 62.6 & .26 & .002 & .002 & 6.7 & 94.80 & 94.74 \\
        9.4 & 60.0 & .23 & .000 & -.001 & 5.1 & 94.68 & 94.74 \\
        \hline
    \end{tabular}
    \caption{Comparison of bias, variance, and coverage of 95\% CI of covariate adjusted pseudovalue regression estimated RMST difference and KM-based RMST difference. Censored, percentage of subjects censored before restriction time $\tau$; Truncated, percentage of subjects who remain at risk of the event at restriction time $\tau$; $r^2$, squared correlation between covariate and pseudovalues; KM bias, bias of KM-based RMST difference; PV bias, bias of covariate adjusted pseudovalue regression estimated RMST difference.}
    \label{tab:linear}
\end{table}

\begin{table}[h]
    \centering
    \begin{tabular}{c c c c c c c c}
        \hline
        Censored & Truncated & $r^2$ & KM & PV & Variance & KM & PV \\
        (\%) & (\%) &  & bias & bias & reduction (\%) & coverage (\%) & coverage (\%) \\
        \hline
        0.0 & 54.1 & .40 & .019 & .019 & 14.5 & 94.80 & 94.62 \\
        0.0 & 53.0 & .34 & .006 & .006 & 10.1 & 94.96 & 95.02 \\
        0.0 & 52.4 & .30 & .001 & .002 & 7.7 & 95.02 & 95.04 \\
        0.0 & 70.0 & .33 & .009 & .009 & 9.5 & 95.58 & 95.44 \\
        0.0 & 68.4 & .27 & .001 & .001 & 6.3 & 95.00 & 94.78 \\
        0.0 & 67.4 & .24 & .000 & .000 & 4.8 & 94.80 & 94.94 \\
        8.2 & 47.9 & .40 & .020 & .020 & 14.1 & 94.88 & 94.88 \\
        11.1 & 45.2 & .33 & .006 & .006 & 9.7 & 95.06 & 95.20 \\
        13.8 & 43.2 & .29 & -.001 & -.001 & 7.3 & 95.32 & 95.14 \\
        4.7 & 66.3 & .32 & .008 & .008 & 9.3 & 95.12 & 94.88 \\
        7.3 & 62.2 & .27 & .002 & .002 & 6.2 & 94.94 & 94.54 \\
        9.5 & 59.7 & .23 & .000 & .000 & 4.6 & 95.06 & 95.18 \\
        \hline
    \end{tabular}
    \caption{Comparison of bias, variance, coverage of 95\% CI of covariate adjusted pseudovalue regression estimated RMST difference and KM-based RMST difference. Covariate with random noise. Censored, percentage of subjects censored before restriction time $\tau$; Truncated, percentage of subjects who remain at risk of the event at restriction time $\tau$; $r^2$, squared correlation between latent covariate (without random errors) and pseudovalues; KM bias, bias of KM-based RMST difference; PV bias, bias of covariate adjusted pseudovalue regression estimated RMST difference.}
    \label{tab:randomx}
\end{table}

\begin{table}[h]
    \centering
    \begin{tabular}{c c c}
        \hline
        & Estimated RMST difference (se) & Variance reduction \\
        \hline
        KM-based & $-6.88 \, (5.19)$ & - \\
        Covariate adjusted pseudovalue regression & $-5.31 \, (4.76)$ & $16.1\%$ \\
        \hline
    \end{tabular}
    \caption{Estimated effect of ceftriaxone infusion on RMST in the ceftriaxone ALS study.}
    \label{tab:case study}
\end{table}

\clearpage
\appendix
\section{Appendices}
\subsection{Distribution of the bias with a random covariate}
\label{sec:A1}

Suppose that $U$ is the matrix consisting of the true covariate, where each row of $U$ is $(1, \mathbbm{1}_{Ti}, u_i)$, and $X$ is the matrix consisting of the observed covariate with random error, where each row of $X$ is $(1, \mathbbm{1}_{Ti}, c_i)$, and $c_i=u_i+\delta_i$, $\mathrm{Var}[\delta]=\sigma^2_{\delta}$. The errors $\epsilon$ and $\delta$ are independent. Let $D$ be a diagonal matrix with the diagonal entries $(0,0,\sigma^2_{\delta})$. By Theorem 4.1 in \citep{davies1975effect}, $b \xrightarrow[]{p} n(U'U+nD)^{-1}D\beta^{\circ}$ as $n \to \infty$. Some algebra shows that the limit of $b_1$ can be expressed as
\begin{equation*}
    \frac{\beta_2^\circ}{\frac{\mathrm{Var}[u]}{\sigma^2_{\delta}(\bar{u}_C-\bar{u}_T)}+\frac{1}{\bar{u}_C+\bar{u}_T}-\frac{\pi(1-\pi)(\bar{u}_C-\bar{u}_T)}{\sigma^2_{\delta}}}.
\end{equation*}

Here, $\bar{u}_T=\sum_i \mathbbm{1}_{T_i}u_i/(n\pi)$, and likewise, $\bar{u}_C$ is the mean of $u_i$ in the control arm. Since the assignment of treatment arm is randomized and independent from the covariate, $\bar{u}_C-\bar{u}_T$ follows a normal distribution with parameters $(0,\mathrm{Var}[u]/(n\pi(1-\pi)))$. Rescale $\bar{u}_C-\bar{u}_T$ to a standard normal distribution, and rescale $\beta_2^{\circ}$ by $\sqrt{\mathrm{Var}[u]}$, Equation \ref{eq:3} is obtained.

As $n \to \infty$, $Z^2/n \to 0$ almost surely, so the $1$st term in the denominator of Equation \ref{eq:3} converges to $\infty$ almost surely, and the $2$nd term in the denominator of Equation \ref{eq:3} converges to $0$ almost surely. Then the denominator converges to $\infty$, and Theorem \ref{claim:2} is proved.

\subsection{Distribution of the variance of treatment effect estimate with a random covariate}
\label{sec:A2}

By Theorem 4.2 and Equation 4.2 in \citep{davies1975effect}, the asymptotic variance of $\hat{\beta}_{\Delta}$ can be written as $\frac{1}{n}\{(\frac{1}{n}U'U+D)^{-1}(\sigma^2_{\epsilon}+\beta^{\circ}{'}\Delta\beta^{\circ})\}$, where $\Delta=(0,0,\sigma^2_{\delta})$. Some algebra shows that the asymptotic variance corresponding to $\hat{\beta}_{\Delta 1}$ is
\begin{equation*}
    \frac{(\sigma^2_{\epsilon}+\beta^{\circ 2}_2\sigma^2_{\delta})(\mathrm{Var}[u]+\sigma^2_{\delta})}{n\pi(1-\pi)(\mathrm{Var}[u]+\sigma^2_{\delta}-\mathrm{Var}[u]Z^2/n)},
\end{equation*}
where $Z$ is a standard normal distributed variable. Rescale $\beta^{\circ}_2$ by $\sqrt{\mathrm{Var}[u]}$ and take the limit, Theorems \ref{claim:3} and \ref{claim:4} are obtained.

\subsection{Additional simulations}
\label{sec:A3}

\begin{figure}[ht]
\caption{Variance reduction of pseudovalue regression estimated RMST difference compared to KM-based RMST difference, versus squared correlation between covariate and pseudovalues. Expectation of latent event times is a quadratic function of the covariate. Grey line is the 45 degree line. The different symbols denote different restriction times, and the different colors denote different distributions for the censoring time.}
\label{fig:quadratic}
\includegraphics[width=10cm]{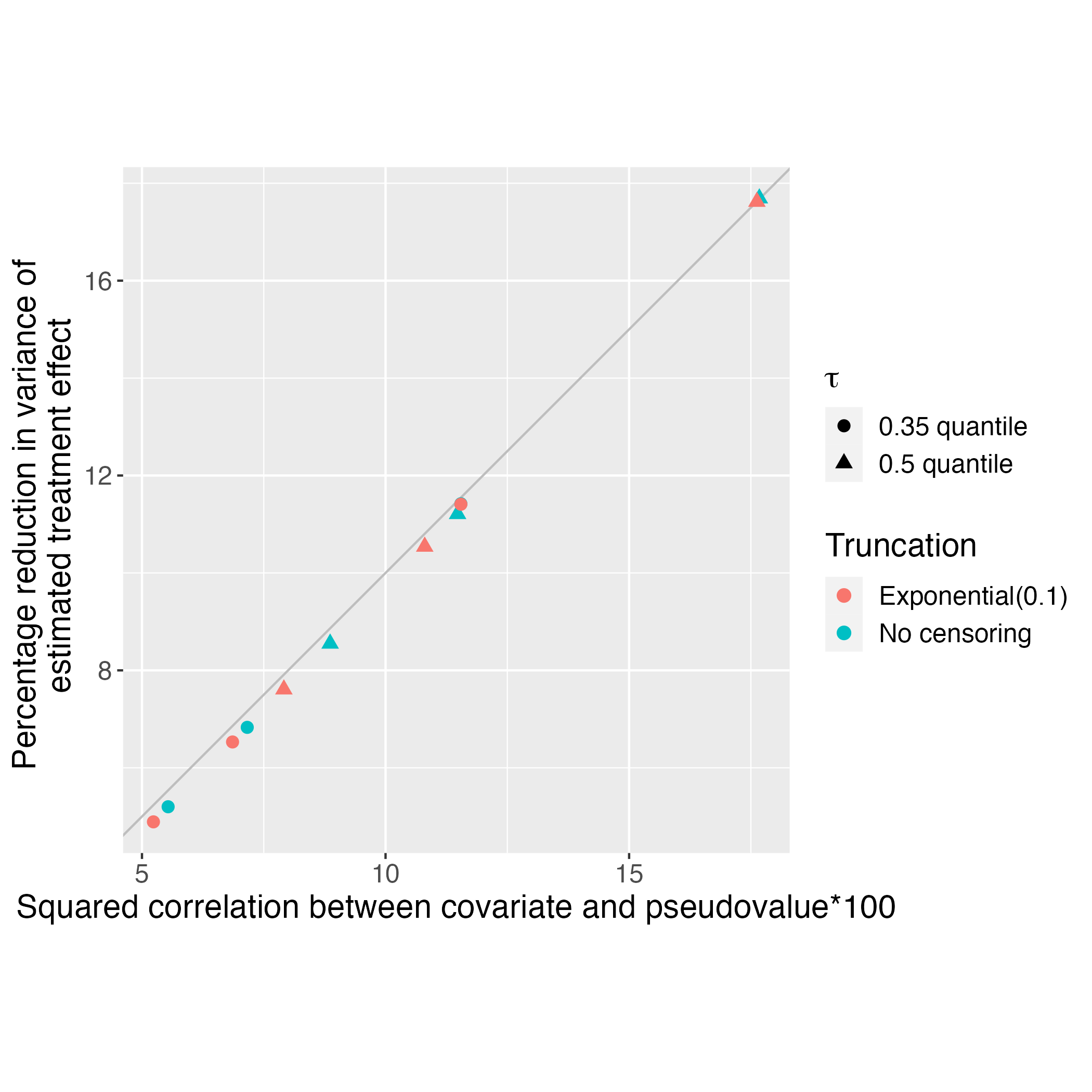}
\centering
\end{figure}

Table \ref{tab:quadratic} and Figure \ref{fig:quadratic} follow the exact same simulation settings as \ref{tab:linear} and Figure \ref{fig:linear}, except that latent event times $Y^{\circ}$ follow an exponential distribution of rate $1/(a+0.5*\mathbbm{1}_T+u^2+3u)$, so that the mean survival time (if there is no censoring) is a quadratic function of the covariate. The constant $a$ takes the value $0$, $1$ or $2$ so that the correlation between the covariate and the response changes.

\begin{table}[t]
    \centering
    \begin{tabular}{c c c c c c c c}
        \hline
        Censored & Truncated & $r^2$ & KM & PV & Variance & KM & PV \\
        (\%) & (\%) &  & bias & bias & reduction (\%) & coverage (\%) & coverage (\%) \\
        \hline
        0.0 & 58.8 & .42 & .018 & .018 & 17.7 & 95.12 & 94.96 \\
        0.0 & 56.5 & .34 & .002 & .002 & 11.2 & 95.08 & 94.88 \\
        0.0 & 54.8 & .30 & .000 & .000 & 8.6 & 95.46 & 95.06 \\
        0.0 & 72.8 & .34 & .011 & .011 & 11.4 & 94.72 & 94.94 \\
        0.0 & 70.5 & .27 & .000 & .000 & 6.8 & 95.04 & 95.14 \\
        0.0 & 69.2 & .24 & .001 & .001 & 5.2 & 94.60 & 94.46 \\
        8.6 & 51.2 & .42 & .028 & .028 & 17.6 & 94.54 & 94.52 \\
        14.4 & 46.1 & .33 & .005 & .005 & 10.5 & 94.70 & 95.02 \\
        19.0 & 42.6 & .28 & .000 & .000 & 7.6 & 95.30 & 94.60 \\
        4.9 & 68.3 & .34 & .012 & .012 & 11.4 & 94.96 & 94.56 \\
        9.5 & 62.8 & .26 & .000 & .000 & 6.5 & 95.24 & 95.02 \\
        13.5 & 58.2 & .23 & .000 & .001 & 4.9 & 94.88 & 94.88 \\
        \hline
    \end{tabular}
    \caption{Comparison of bias, variance, coverage of 95\% CI of covariate adjusted pseudovalue regression estimated RMST difference and KM-based RMST difference. Expectation of latent event times is a quadratic function of the covariate. Censored, percentage of subjects censored before restriction time $\tau$; Truncated, percentage of subjects who remain at risk of the event at restriction time $\tau$; $r^2$, squared correlation between covariate and pseudovalues; KM bias, bias of KM-based RMST difference; PV bias, bias of covariate adjusted pseudovalue regression estimated RMST difference.}
    \label{tab:quadratic}
\end{table}

\end{document}